# Fastest Distributed Consensus on Petal Networks

Saber Jafarizadeh
Department of Electrical Engineering
Sharif University of Technology, Azadi Ave, Tehran, Iran
Email: jafarizadeh@ee.sharif.edu

*Abstract*— **Providing an analytical solution for the problem of finding Fastest Distributed Consensus (FDC) is one of the challenging problems in the field of sensor networks. Here in this work we present analytical solution for the problem of fastest distributed consensus averaging algorithm by means of stratification and semi-definite programming, for two particular types of Petal networks, namely symmetric and Complete Cored Symmetric (CCS) Petal networks. Our method in this paper is based on convexity of fastest distributed consensus averaging problem, and inductive comparing of the characteristic polynomials initiated by slackness conditions in order to find the optimal weights. Also certain types of leaves are introduced along with their optimal weights which are not achievable by the method used in this work if these leaves are considered individually.**

*Index Terms*— **Distributed Consensus, Weight Optimization, Semidefinite Programming, Sensor Networks.**

## I. INTRODUCTION

Distributed consensus has appeared as one of the most important and primary problems in the context of distributed computation and it has received renewed interest in the field of sensor networks, where some of its applications include distributed agreement, synchronization problems, [1] and load balancing in parallel computers [2]. Solving fastest distributed consensus averaging problem over networks with different topologies is one of the primary and challenging problems in this issue. Most of the methods proposed so far deal with the FDC averaging algorithm problem by numerical convex optimization methods and in general no closed-form solution for finding FDC has been offered up to now, except the previous works by author and [3].

In [4], the author proposes an analytical solution for FDC problem based on Stratification and Semidefinite Programming (SDP), by imposing the slackness conditions and comparing the corresponding characteristic polynomials of blocks of weight matrix for two networks, which are special cases of Petal networks and in [5] the author has solved FDC problem analytically over branches connected to arbitrary networks where it has been proved that the obtained optimal weights are independent of the rest of the network.

Following the same procedure used in [4, 5] we have managed to solve FDC problem over two particular types of Petal networks, namely symmetric and CCS Petal networks, in addition we have introduced certain types of leaves which can be used in symmetric and CCS Petal configurations along with their optimal weights where these optimal weights are not achievable if we consider the leaves individually.

The organization of the paper is as follows. Section II is an overview of the materials used in development of the paper, including relevant concepts from distributed consensus averaging algorithm, graph symmetry and semidefinite programming. In section III we present the symmetric and complete cored symmetric Petal (CCS) networks along with other kinds of leaves which can be used in symmetric Petal and CCS Petal configurations with the obtained optimal weights and the corresponding Second Largest Eigenvalue Modulus (*SLEM*). Section IV is devoted to proof of main results of paper for symmetric and CCS Petal networks and symmetric and CCS Petal networks with $G_{m,k}$ Leaves and section V presents the conclusion and topics for future research.

## II. Preliminaries

This section introduces the notation used in the paper and reviews relevant concepts from distributed consensus averaging algorithm, graph symmetry and semidefinite programming.

### *A. Distributed Consensus*

We consider a network $\mathcal{N}$ with the associated graph $\mathcal{G} = (\mathcal{V}, \mathcal{E})$ consisting of a set of nodes $\mathcal{V}$ and a set of edges $\mathcal{E}$ where each edge $\{i, j\} \in \mathcal{E}$ is an unordered pair of distinct nodes.

Each node $i$ holds an initial scalar value $x_i(0) \in \mathbf{R}$, and $x^T(0) = (x_1(0), \dots, x_n(0))$ denotes the vector of initial node values on the network. Within the network two nodes can communicate with each other, if and only if they are neighbors.

The main purpose of distributed consensus averaging is to compute the average of the initial node values, $(1/n)\sum_{i=1}^{n} x_i(0)$ via a distributed algorithm, in which the nodes only communicate with their neighbors.

In this work, we consider distributed linear iterations, which have the form

$$x_i(t+1) = W_{ii}x_i(t) + \sum_{j \neq i} W_{ij}x_j(t), \quad i = 1, \dots, n$$

where $t = 0,1,2, \dots$ is the discrete time index and $W_{ij}$ is the weight on $x_j$ at node $i$ and the weight matrix have the same sparsity pattern as the adjacency matrix of the network's associated graph or $W_{ij} = 0$ if $\{i, j\} \notin \mathcal{E}$, this iteration can be written in vector form as

$$x(t+1) = Wx(t)$$

The linear iteration (1) implies that $x(t) = W^t x(0)$ for $= 0,1,2,\ldots$ . We want to choose the weight matrix $W$ so that for any initial value $x(0),\ x(t)$ converges to the average vector $\bar{x} = (\mathbf{1}^T x(0)/n)\mathbf{1} = (\mathbf{1}\mathbf{1}^T/n)x(0)$ i.e.

$$\lim_{t\to\infty} x(t) = \lim_{t\to\infty} W^t x(0) = \frac{\mathbf{1}\mathbf{1}^T}{n} x(0)$$

(Here $\mathbf{1}$ denotes the column vector with all coefficients one). This is equivalent to the matrix equation

$$\lim_{t\to\infty} W^t = \frac{\mathbf{1}\mathbf{1}^T}{n} \tag{1}$$

Assuming (1) holds, the *convergence factor* can be defined as

$$r(W) = sup \frac{\|x(t+1) - \bar{x}\|_2}{\|x(t) - \bar{x}\|_2}$$

where $\|\cdot\|_2$ denotes the spectral norm, or maximum singular value. The FDC problem in terms of the convergence factor can be expressed as the following optimization problem:

$$\begin{aligned} \min_{W} \quad & r(W) \\ s.t. \quad & \lim_{t\to\infty} W^t = \mathbf{1}\mathbf{1}^T/n, \\ & \forall\{i,j\} \notin \mathcal{E}: W_{ij} = 0 \end{aligned} \tag{2}$$

where $W$ is the optimization variable, and the network is the problem data.

The FDC problem (2) is closely related to the problem of finding the fastest mixing Markov chain on a graph [3]; the only difference in the two problem formulations is that in the FDC problem, the weights can be (and the optimal ones often are) negative, hence faster convergence could be achieved compared with the fastest mixing Markov chain on the same graph.

In [6] it has been shown that the necessary and sufficient conditions for the matrix equation (1) to hold is that one is a simple eigenvalue of $W$ associated with the eigenvector $\mathbf{1}$, and all other eigenvalues are strictly less than one in magnitude. Moreover in [6] FDC problem has been formulated as the following minimization problem

$$\begin{aligned} \min_{W} \quad & \max(\lambda_2, -\lambda_n) \\ s.t. \quad & W = W^T, W\mathbf{1} = \mathbf{1} \\ & \forall\{i,j\} \notin \mathcal{E}: W_{ij} = 0 \end{aligned}$$

where $1 = \lambda_1 \geq \lambda_2 \geq \cdots \geq \lambda_n \geq -1$ are eigenvalues of $W$ arranged in decreasing order and $\max(\lambda_2, -\lambda_n)$ is the *Second Largest Eigenvalue Modulus* (*SLEM*) of $W$, and the main problem can be derived in the semidefinite programming form as [6]:

$$\begin{aligned} \min_{W} \quad & s \\ s.t. \quad & -sI \preccurlyeq W - \mathbf{1}\mathbf{1}^T/n \preccurlyeq sI \\ & W = W^T, W\mathbf{1} = \mathbf{1} \end{aligned} \tag{3}$$

$$\forall \{i,j\} \notin \mathcal{E}: W_{ij} = 0$$

We refer to problem (3) as the Fastest Distributed Consensus (FDC) averaging problem.

*B. Symmetry of Graphs*

An automorphism of a graph $\mathcal{G} = (\mathcal{V}, \mathcal{E})$ is a permutation $\sigma$ of $\mathcal{V}$ such that $\{i,j\} \in \mathcal{E}$ if and only if $\{\sigma(i), \sigma(j)\} \in \mathcal{E}$, the set of all such permutations, with composition as the group operation, is called the automorphism group of the graph and denoted by $Aut(\mathcal{G})$. For a vertex $i \in \mathcal{V}$, the set of all images $\sigma(i)$, as $\sigma$ varies through a subgroup $G \subseteq Aut(\mathcal{G})$, is called the orbit of $i$ under the action of $G$. The vertex set $\mathcal{V}$ can be written as disjoint union of distinct orbits. In [7], it has been shown that the weights on the edges within an orbit must be the same.

*C. Semidefinite Programming (SDP)*

SDP is a particular type of convex optimization problem [8]. An SDP problem requires minimizing a linear function subject to a linear matrix inequality (LMI) constraint [9]:

$$\min \quad \rho = c^T x,$$

$$s.t. \quad F(x) \geq 0$$

where $c$ is a given vector, $x^T = (x_1, \dots, x_n)$, and $F(x) = F_0 + \sum_i x_i F_i$, for some fixed Hermitian matrices $F_i$. The inequality sign in $F(x) \geq 0$ means that $F(x)$ is positive semidefinite.

This problem is called the primal problem. Vectors $x$ whose components are the variables of the problem and satisfy the constraint $F(x) \geq 0$ are called primal feasible points, and if they satisfy $F(x) > 0$, they are called strictly feasible points. The minimal objective value $c^T x$ is by convention denoted by $\rho^*$ and is called the primal optimal value.

Due to the convexity of the set of feasible points, SDP has a nice duality structure, with the associated dual program being:

$$\max \quad -Tr[F_0 Z]$$

$$s.t. \quad Z \geq 0$$

$$Tr[F_i Z] = c_i$$

Here the variable is the real symmetric (or Hermitian) positive matrix $Z$, and the data $c$, $F_i$ are the same as in the primal problem. Correspondingly, matrix $Z$ satisfying the constraints is called dual feasible (or strictly dual feasible if $Z > 0$). The maximal objective value of $-Tr[F_0 Z]$, i.e. the dual optimal value is denoted by $d^*$.

The objective value of a primal (dual) feasible point is an upper (lower) bound on $\rho^*(d^*)$. The main reason why one is interested in the dual problem is that one can prove that $d^* \leq \rho^*$, and under relatively mild assumptions, we can have $\rho^* = d^*$. If the equality holds, one can prove the following optimality condition on $x$.

A primal feasible $x$ and a dual feasible $Z$ are optimal, which is denoted by $\hat{x}$ and $\hat{Z}$, if and only if

$$F(\hat{x})\hat{Z} = \hat{Z}F(\hat{x}) = 0. \quad (4)$$

This latter condition is called the complementary slackness condition.

In one way or another, numerical methods for solving SDP problems always exploit the inequality $d \le d^* \le \rho^* \le \rho$, where $d$ and $\rho$ are the objective values for any dual feasible point and primal feasible point, respectively. The difference

$$\rho^* - d^* = c^T x + Tr[F_0 Z] = Tr[F(x)Z] \ge 0$$

is called the duality gap. If the equality $d^* = \rho^*$ holds, i.e. the optimal duality gap is zero, and then we say that strong duality holds.

## III. MAIN RESULTS

This section presents the main results of the paper. Here we introduce the symmetric Petal and complete cored symmetric (CCS) Petal networks along with other kinds of leaves which can be used in symmetric Petal and CCS Petal configurations with the obtained optimal weights and the corresponding *SLEM.*

### *A. Symmetric Petal Network*

A symmetric Petal network of order $(n, m, k)$ consists of $n$ similar leaf graphs of order $(m, k)$ which share one of their ending nodes together and each leaf graph of order $(m, k)$ contains $k$ path graphs of length $m$ which are connected to each other from both ending nodes. A symmetric petal network of order $(n = 5, m = 3, k = 4)$ is depicted in Fig.1.

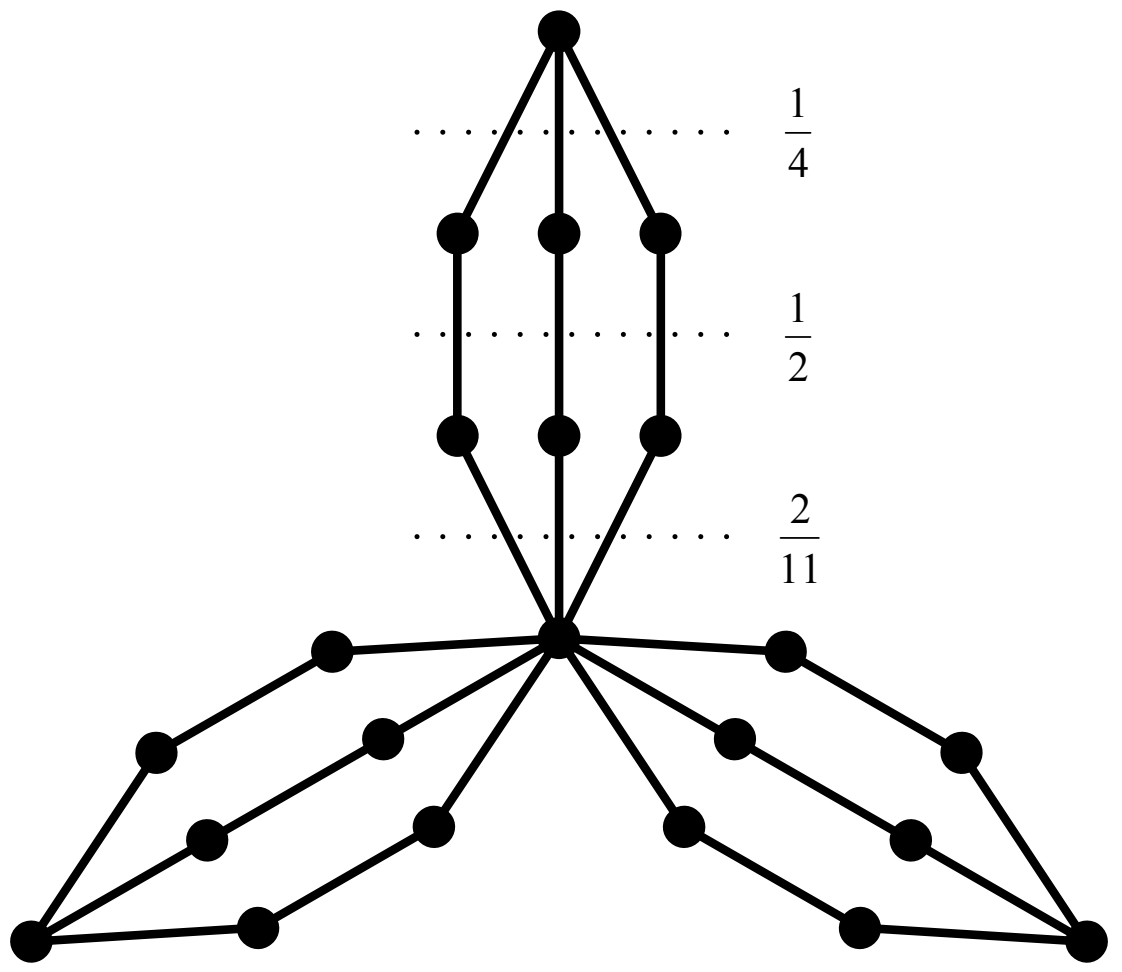

Fig.1. A symmetric Petal network of order $(n = 3, m = 4, k = 3)$

The optimal weights over the edges of a symmetric petal network equals $1/2$ except the edges connected to central node and the edges connected to last node (with maximum distance from central node) of each leaf where the optimal weights over these edges equals $2/(2 + nk)$ and $1/(1 + k)$, respectively (as weighted in Fig.1. for one leaf). The *SLEM* of the network equals $\cos(\theta)$ where $\theta$ is the smallest root of $(nk - 2)\big(\sin\big((m - 1)\theta\big) + \sin\big((m + 1)\theta\big)\big) + 2nk \sin(m\theta) = 0$ between $0$ and $\pi$. The *SLEM* of symmetric Petal network of order $(n, m, k)$ for different numbers of $n, m$ and $k$ has been listed below in Table 1.

| $(n,m,k)$ | *SLEM* | $(n,m,k)$ | *SLEM* |
|---|---|---|---|
| (2,2,1) | 0.80901 | (3,2,1) | 0.83851 |
| (2,2,2) | 0.80473 | (3,2,2) | 0.84824 |
| (2,2,3) | 0.82569 | (3,2,3) | 0.87040 |
| (2,3,1) | 0.90096 | (3,3,1) | 0.91294 |
| (2,3,2) | 0.89987 | (3,3,2) | 0.91935 |
| (2,3,3) | 0.91143 | (3,3,3) | 0.93210 |
| (2,3,4) | 0.92278 | (4,3,5) | 0.96107 |

Table. 1. *SLEM* of symmetric Petal network of order $(n,m,k)$ for different numbers of $n,m$ and $k$.

Symmetric Petal network for $k=1$ reduces to symmetric star network where the obtained results are in agreement with those of [4]. The leaf of symmetric Petal network for $k=2$ reduces to a cycle network where the optimal weights of FDC problem over cycle network [7] depend on global topology of network while here the weights of the edges except for those connected to central node are independent of length of leaves. As it is obvious from the results given in Table.1. *SLEM* of a symmetric Petal network increases by number of path graphs on each leaf due to increase in number of nodes except for $n=k=2$ since in the case of $n=k=2$ by increasing the number of path graphs on each leaf from 1 to 2 the bottleneck effect of middle nodes of path graphs on leaves is reduced but after $k=3$ the bottleneck effect of the central node is dominant and in the case of $n>2$, *SLEM* of a symmetric Petal network increases by number of path graphs on each leaf due to increase in number of nodes.

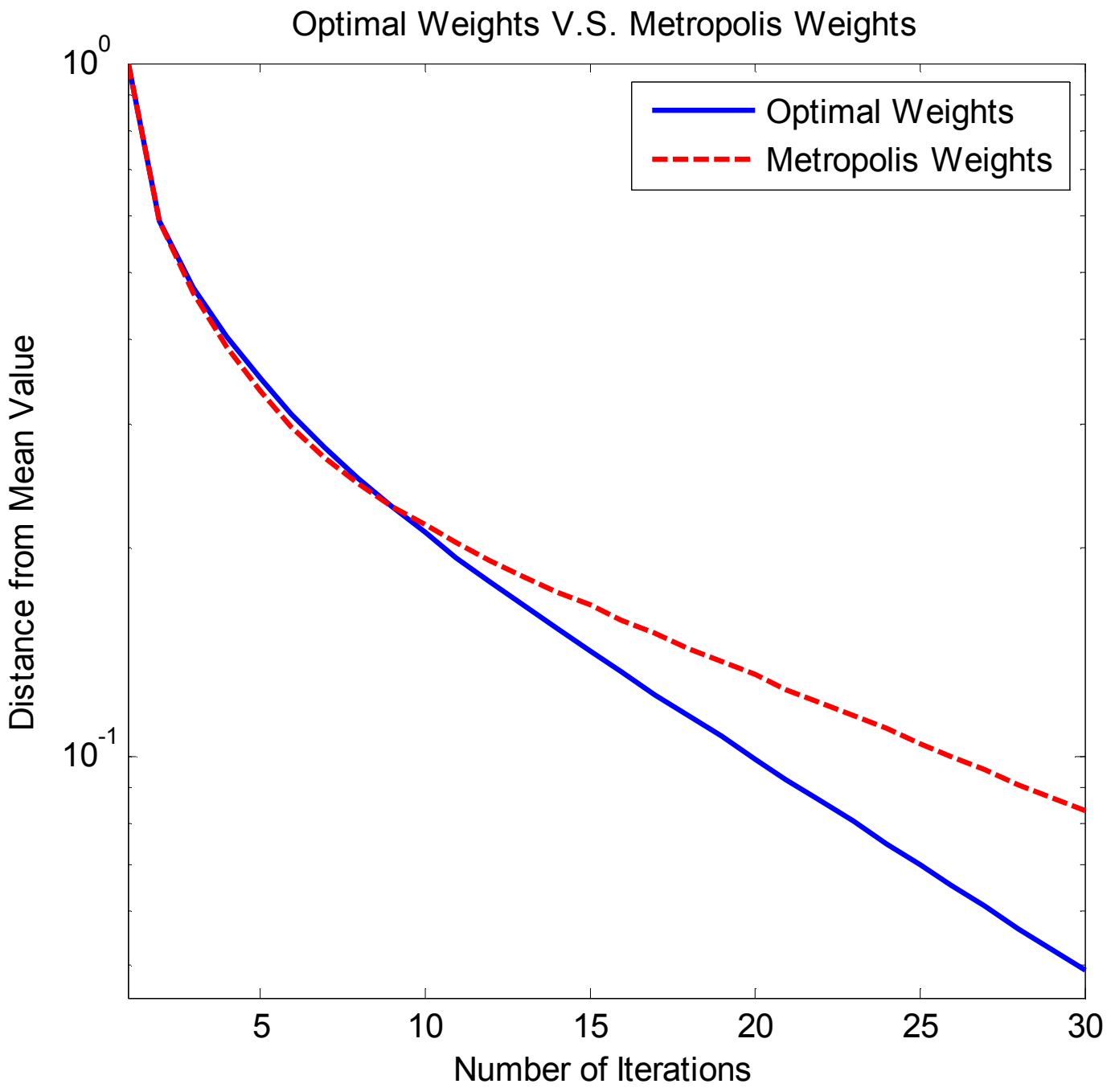

Fig.2. Normalized Euclidean Distance of vector of node values from the stationary distribution in terms of number of iterations.

The only similarity between the obtained optimal weights and those obtained from Metropolis-Hasting method is over the middle edges of path graphs within leaves where their weights equal $1/2$. We have compared the obtained optimal weights and those obtained from Metropolis-Hasting method over symmetric Petal network depicted in Fig.2. in a per step manner.

In Fig.2. normalized Euclidean distance of vector of node values from the stationary distribution in terms of number of iterations for the symmetric Petal network depicted in Fig.1. is presented

As it is obvious from Fig.2. at first few iterations Metropolis-Hasting method and the optimal weights have the same mixing rate per step but after first few iterations optimal weights achieve better mixing rate per step since due to their smaller *SLEM*.

*B. Complete Cored Symmetric Petal (CCS Petal) Network*

A CCS Petal network of order $(n, m, k)$ consists of $n$ similar leaf graphs of order $(m, k)$ where the leaves are connected to each other at one end such that the connected part of leaves form a complete graph. A CCS Petal network of order $(n = 3, m = 4, k=3$ is depicted in Fig.3.

The optimal weights over the edges of the central complete part equals $1/n$ and the rest of the edges equals $1/2$ except the edges connected to central complete part and the edges connected to last node (with maximum distance from central complete part) of each leaf where the optimal weights over these edges equals $1/(1 + k)$ (as weighted in Fig.3. for one leaf).

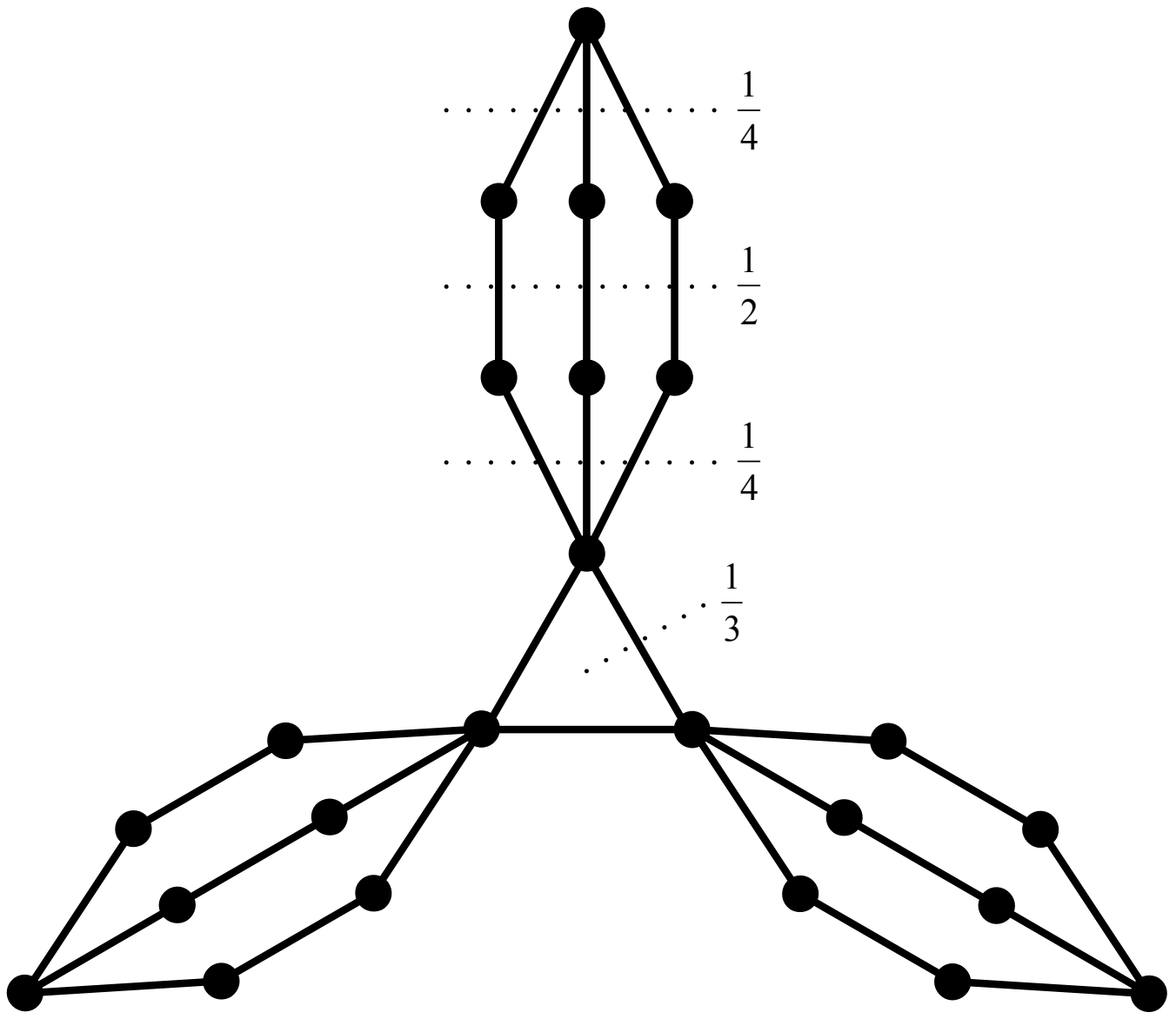

Fig.3. A complete cored symmetric Petal network of order $(n = 3, m = 4, k = 3)$

The *SLEM* of CCS Petal network of order $(n, m, k)$ for different numbers of $n, m$ and $k$ has been listed below in Table 2.

| $(n, m, k)$ | *SLEM* | $(n, m, k)$ | *SLEM* |
|---|---|---|---|
| (2,2,1) | 0.86602 | (3,2,1) | 0.86602 |
| (2,2,2) | 0.88191 | (3,2,2) | 0.88191 |
| (2,2,3) | 0.90138 | (3,2,3) | 0.90138 |
| (2,3,1) | 0.92387 | (3,3,1) | 0.92387 |
| (2,3,2) | 0.93417 | (3,3,2) | 0.93417 |
| (2,3,3) | 0.94619 | (3,3,3) | 0.94619 |
| (2,3,4) | 0.95514 | (4,3,5) | 0.96172 |

Table. 2. *SLEM* of CCS Petal network of order $(n, m, k)$ for different numbers of $n, m$ and $k$.

A CCS Petal network for $k = 1$ reduces to complete cored symmetric star network where the obtained results are in agreement with those of [4]. As it is obvious from the results given in Table.1. *SLEM* of a CCS Petal network does not change with $n$, the number of leaves since the central complete graph acts like a highway and does not make any restrictions over *SLEM* then *SLEM* only depends on topology of each individual leaf and does not change with number of leaves, which is true for other networks with complete cored structures. This a great advantage of networks with complete cored configurations over other similar networks with central node configurations.

Here in the following subsections we introduce different graphs which can be replaced as leaves of symmetric and CCS Petal networks, together with their corresponding optimal weights for symmetric and CCS Petal configurations, respectively.

*C. Symmetric $G_{m,k}$ Leaf*

A symmetric $G_{m,k}$ network consists of two balanced $k$-ary tree of height $m$ with the $k^m$ leaves of the left tree identified with the $k^m$ leaves of the right tree in the simple way as shown in Fig.4. for $m = 3, k = 2$.

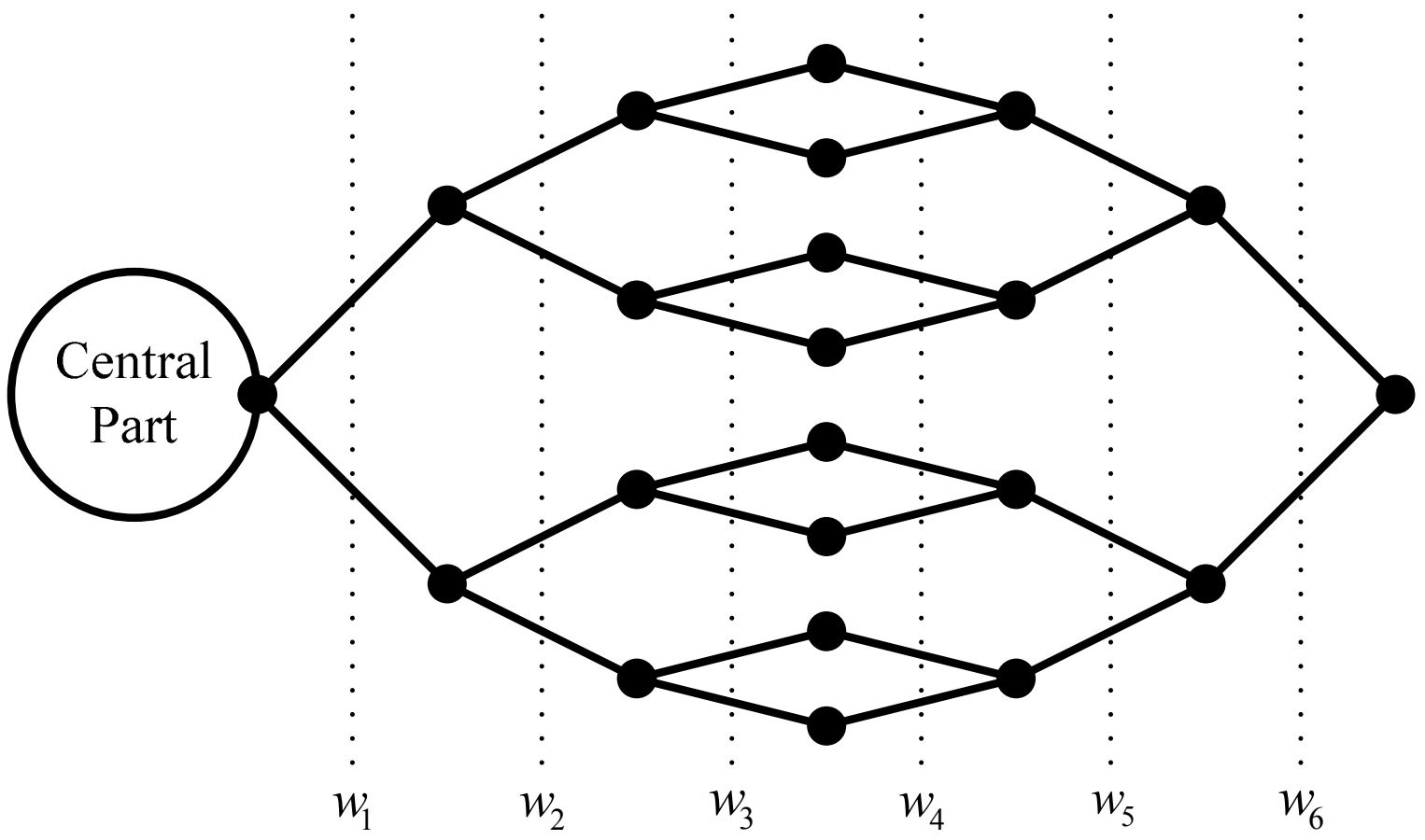

Fig.4. A symmetric $G_{m,k}$ network for $m = 3, k = 2$ as a leaf.

A balanced $k$-ary tree is a tree network of height $m$ where each node except the leave nodes (nodes without children) has $k$ children

In a symmetric petal network with symmetric $G_{m,k}$ leaves where $n$ similar symmetric $G_{m,k}$ networks are connected to a central node, the optimal weights over the edges connected to central node equals $2/(2 + nk)$ and for the rest of edges equals $1/(1 + k)$ and in a CCS petal network with symmetric $G_{m,k}$ leaves where $n$ similar symmetric $G_{m,k}$ networks are connected to each other at one end such that the connected part of leaves form a complete graph, the optimal weights over the edges of central complete graph equals $1/n$ and for the rest of edges equals $1/(1 + k)$.

*D.* Asymmetric $G_{m_1,m_2}$ *Leaf*

An asymmetric $G_{m_1,m_2}$ network is a symmetric $G_{m,k}$ network where every node at $i$-th depth of each tree have $k_i$ children, in such a way that both trees have the same number of leaf nodes. An asymmetric $G_{m_1,m_2}$ leaf is depicted in Fig.5.

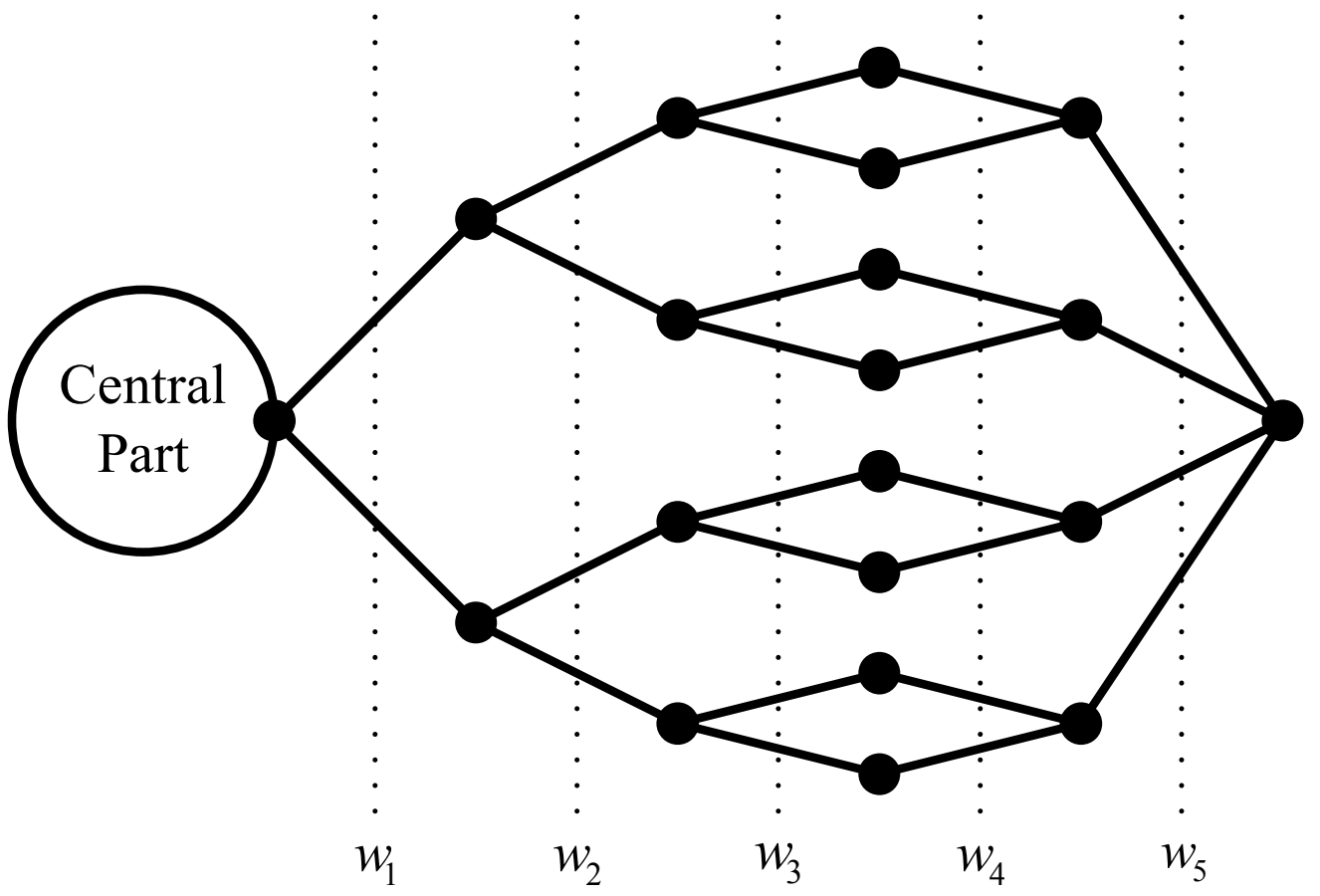

Fig.5. An asymmetric $G_{m_1,m_2}$ as a leaf for $m = 3, m' = 2, k_1 = k_2 = k_3 = k_4 = 2, k_5 = 4$.

In a symmetric petal network with asymmetric $G_{m_1,m_2}$ leaves where $n$ similar asymmetric $G_{m_1,m_2}$ networks are connected to a central node, the optimal weights over the edges connected to central node equals $2/(2 + nk_1)$ and for the of edges on $i$-th depth of each tree network of leaves equals $1/(1 + k_i)$ and in a CCS petal network with asymmetric $G_{m_1,m_2}$ leaves where $n$ similar asymmetric $G_{m_1,m_2}$ networks are connected to each other at one end such that the connected part of leaves form a complete graph, the optimal weights over the edges of central complete graph equals $1/n$ and for the edges on $i$-th depth of each tree network of leaves equals $1/(1 + k_i)$.

*E. Extended Leaves*

For the leaves built from combination of the leaves introduced in previous sections, the obtained optimal weights still hold true, in both symmetric Petal and CCS Petal configurations. In Fig.6. a combined leaf is depicted.

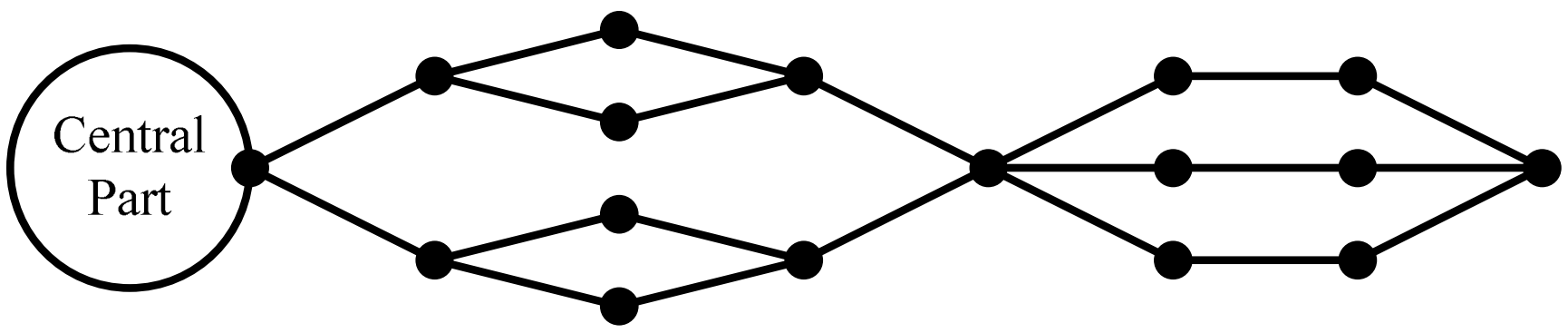

Fig.6. A combined leaf.

Also in [5, 7] five different branches or leaves are introduced, namely, path, Lollipop, Semi-Complete, Ladder and tree branches, where these branches can be added to leaves introduced in this paper while their optimal weights remain unchanged.

## IV. Proof of Main Results

In this section solution of fastest distributed consensus averaging problem and determination of optimal weights over symmetric and complete cored symmetric (CCS) Petal networks along with other kinds of leaves introduced in section III are presented.

### *A. Symmetric Petal Network*

Here we consider a symmetric Petal network of order $n, m, k$ with the undirected associated connectivity graph $\mathcal{G} = (\mathcal{V}, \mathcal{E})$ consisting of $|\mathcal{V}| = 1 + n(k(m-1)+1)$ nodes and $|\mathcal{E}| = nkm$ edges, where the set of nodes is denoted by $\mathcal{V} = \{(0,0,0), (i,j,q)\}$ where $i, j$ and $k$ vary from 1 to $m$, 1 to $n$ and 1 to $k$ respectively. (see Fig.1 for $n = 3, m = 4, k = 3$).

Automorphism of symmetric Petal network is $S_k$ permutation of nodes of each Leaf on $i$-th distance from central node for $i = 1, \dots, m$ and $S_n$ permutation of leaves hence according to subsection II-B it has $m+1$ class of edge orbits and it suffices to consider just $m$ weights $w_1, w_2, \dots, w_m$ (as labeled in Fig. 1. for $n = 3, m = 4, k = 3$), and consequently the weight matrix for the network can be defined as

$$W_{(i,j,q),(\mu,\rho,\eta)} = \begin{cases} w_1 \;\; \textbf{for} \;\; i = j = q = 0, \;\; \mu = 1, \rho = 1, \dots, n, \;\; \eta = 1, \dots, k \\ w_1 \;\; \textbf{for} \;\; i = 1, j = 1, \dots, n, q = 1, \dots, k, \;\; \mu = \rho = \eta = 0 \\ w_{i+1} \;\; \textbf{for} \;\; i = 1, \dots, m-1, \mu = i+1, j = \rho = 1, \dots, n, q = \eta = 1, \dots, k \\ w_i \;\; \textbf{for} \;\; i = 2, \dots, m, \mu = i-1, j = \rho = 1, \dots, n, q = \eta = 1, \dots, k \\ 1 - nkw_1 \;\; \textbf{for} \;\; i = j = q = \mu = \rho = \eta = 0 \\ 1 - w_i - w_{i+1} \;\; \textbf{for} \;\; i = \mu = 1, \dots, m-1, \;\; j = \rho = 1, \dots, n, \;\; q = \eta = 1, \dots, k \\ 1 - w_m \;\; \textbf{for} \;\; i = \mu = m \;\; j = \rho = 1, \dots, n, \;\; q = \eta = 1, \dots, k \end{cases}$$

We associate with the node $(i,,j,q)$ the $|\mathcal{V}| \times 1$ column vector $e_{i,j,q} = e_i \otimes e_j \otimes e_q$ for $\{i,j,q\} = \{i = 1, \dots, m, \;\; j = 1, \dots, n, \;\; q = 1, \dots, k\} \cup \{i = j = q = 0\}$ where $e_i, e_j$ and $e_q$ are $(m+1) \times 1, (n+1) \times 1$ and $(q+1) \times 1$ column vectors with one in the $i$-th, $j$-th and $q$-th position respectively and zero elsewhere. Introducing the new basis $\varphi_0 = e_{0,0,0}$ and

$$\varphi_{i,\mu_1,\mu_2} = \frac{1}{\sqrt{nk}} \sum_{j=1}^{n} \omega_1^{(j-1)\mu_1} \sum_{q=1}^{k} \omega_2^{(q-1)\mu_2} e_{i,j,q} \;\; \textbf{for} \;\; i = 1, \dots, m, \;\; \mu_1 = 0, \dots, n-1, \;\; \mu_2 = 0, \dots, k-1$$

with $\omega_1 = e^{j\frac{2\pi}{n}}$ and $\omega_2 = e^{j\frac{2\pi}{k}}$ and using Stratification method [7, 11, 12, 13@4th paper] the weight matrix $W$ for symmetric Petal network in the new basis takes the block diagonal form with diagonal blocks $W_1, W_2$ and $W_3$ defined as:

$$W_1 = \begin{bmatrix} 1 - nkw_1 & \sqrt{nk}w_1 & 0 & & & \\ \sqrt{nk}w_1 & 1 - w_1 - w_2 & w_2 & \ddots & & \\ 0 & w_2 & 1 - w_2 - w_3 & \ddots & 0 & \\ & \ddots & \ddots & \ddots & w_{m-1} & 0 \\ & & 0 & w_{m-1} & 1 - w_{m-1} - w_m & \sqrt{k}w_m \\ & & & 0 & \sqrt{k}w_m & 1 - kw_m \end{bmatrix} \quad \text{(5-a)}$$

$$W_2 = \begin{bmatrix} 1-w_1-w_2 & w_2 & 0 & & \\ w_2 & 1-w_2-w_3 & w_3 & \ddots & \\ 0 & w_3 & \ddots & \ddots & 0 \\ & \ddots & \ddots & 1-w_{m-1}-w_m & \sqrt{k}w_m \\ & & 0 & \sqrt{k}w_m & 1-kw_m \end{bmatrix} \quad \text{(5-b)}$$

$$W_3 = \begin{bmatrix} 1-w_1-w_2 & w_2 & 0 & & \\ w_2 & 1-w_2-w_3 & w_3 & \ddots & \\ 0 & w_3 & \ddots & \ddots & \\ & \ddots & \ddots & 1-w_{m-2}-w_{m-1} & w_{m-1} \\ & & & w_{m-1} & 1-w_{m-1}-w_m \end{bmatrix} \quad \text{(5-c)}$$

Considering the fact that $W_2$ and $W_3$ are sub matrices of $W_1$ and $W_2$, respectively and using *Cauchy Interlacing Theorem*,

*Theorem 1* (*Cauchy Interlacing Theorem*) [15]:

Let A and B be $\mathrm{n} \times \mathrm{n}$ and $\mathrm{m} \times \mathrm{m}$ matrices, where $\mathrm{m} \leq \mathrm{n}$, B is called a compression of A if there exists an orthogonal projection P onto a subspace of dimension m such that $\mathrm{PAP} = \mathrm{B}$. The Cauchy interlacing theorem states that If the eigenvalues of A are $\lambda_1(\mathrm{A}) \leq \cdots \leq \lambda_\mathrm{n}(\mathrm{A})$, and those of B are $\lambda_1(\mathrm{B}) \leq \cdots \leq \lambda_\mathrm{m}(\mathrm{B})$, then for all j,

$$\lambda_\mathrm{j}(\mathrm{A}) \leq \lambda_\mathrm{j}(\mathrm{B}) \leq \lambda_{\mathrm{n-m+j}}(\mathrm{A})$$

Notice that, when $n-m=1$, we have

$$\lambda_j(A) \leq \lambda_j(B) \leq \lambda_{j+1}(A)$$

we can state the following corollary for the elements of $W_0'$ ,

In the case of $n=1$, after stratification the weight matrix $W_2$ does not exist and consequently Cauchy interlacing theorem will not be true thus the followings are true for $n \geq 2$.

*Corollary 1*,

For $W_1, W_2$ and $W_3$ given in (5), theorem 1 implies the following relations between the eigenvalues of $W_1, W_2$ and $W_3$

$$\lambda_{|\mathcal{V}|}(W) = \lambda_{m+1}(W_1) \leq \lambda_m(W_2) \leq \cdots \leq \lambda_2(W_1) \leq \lambda_1(W_2) \leq \lambda_1(W_1) = 1 \quad \text{(6-a)}$$

$$\lambda_m(W_2) \leq \lambda_{m-1}(W_3) \leq \cdots \leq \lambda_2(W_2) \leq \lambda_1(W_3) \leq \lambda_1(W_2) \quad \text{(6-b)}$$

It is obvious from above relations that $\lambda_2(W)$ and $\lambda_{|\mathcal{V}|}(W)$ are amongst the eigenvalues of $W_1$and $W_2$, respectively.

Based on corollary 1 and subsection II-A, one can express FDC problem for symmetric Petal network in the form of semidefinite programming as:

$$\begin{aligned} \min \quad & s \\ s.t. \quad & -sI \leq W_1 - \boldsymbol{v}\boldsymbol{v}^T \\ & W_2 \leq sI \end{aligned} \quad (7)$$

where $\boldsymbol{v}$ is a $(m+1) \times 1$ column vector defined as:

$$\boldsymbol{v}(i) = \frac{1}{n + nk(m-1) + 1} \times \begin{cases} 1 \ \ \textbf{for} \ \ i = 1, \\ \sqrt{nk} \ \ \textbf{for} \ \ i = 2, \dots, m, \\ \sqrt{n} \ \ \textbf{for} \ \ i = m+1, \end{cases}$$

which is eigenvector of $W_1$ corresponding to the eigenvalue one. The matrices $W_1$ and $W_2$ can be written as

$$W_1 = I_{m+1} - \sum_{i=1}^{m} w_i \, \boldsymbol{\alpha}_i \boldsymbol{\alpha}_i^T \tag{8-a}$$

$$W_2 = I_m - \sum_{i=1}^{m} w_i \boldsymbol{\beta}_i \boldsymbol{\beta}_i^T \tag{8-b}$$

where $\boldsymbol{\alpha}_i$ and $\boldsymbol{\beta}_i$are $(m+1) \times 1$ and $m \times 1$ column vectors, respectively defined as:

$$\boldsymbol{\alpha}_1(j) = \begin{cases} \sqrt{nk} \ \ j = 1 \\ -1 \ \ j = 2 \\ 0 \ \ Otherwise \end{cases}$$

$$\boldsymbol{\alpha}_i(j) = \begin{cases} 1 \ \ j = i \\ -1 \ \ j = i+1 \\ 0 \ \ Otherwise \end{cases} \ \textbf{for} \ \ i = 2, \dots, m-1,$$

$$\boldsymbol{\alpha}_m(j) = \begin{cases} 1 \quad j = m \\ -\sqrt{k} \ \ j = i+1 \\ 0 \ \ Otherwise \end{cases}$$

$$\boldsymbol{\beta}_1(j) = \begin{cases} -1 \ \ j = 1 \\ 0 \ \ Otherwise \end{cases}$$

$$\boldsymbol{\beta}_i(j) = \begin{cases} 1 \ \ j = i-1 \\ -1 \ \ j = i \\ 0 \ \ Otherwise \end{cases} \qquad \textbf{for} \ \ i = 2, \dots, m-1.$$

$$\boldsymbol{\beta}_m(j) = \begin{cases} 1 \ \ j = m-1 \\ -\sqrt{k} \ \ j = m \\ 0 \ \ Otherwise \end{cases}$$

In order to formulate problem (7) in the form of standard semidefinite programming described in section II-C, we define $F_i, c_i$ and $x$ as below:

$$F_0 = \begin{bmatrix} -I_m & 0 \\ 0 & I_{m+1} - \boldsymbol{v}\boldsymbol{v}^T \end{bmatrix}$$

$$F_i = \begin{bmatrix} \boldsymbol{\beta}_i \boldsymbol{\beta}_i^T & 0 \\ 0 & -\boldsymbol{\alpha}_i \boldsymbol{\alpha}_i^T \end{bmatrix} \ \textbf{for} \ \ i = 1, \dots, m,$$

$$F_{m+1} = I_{2m+1}$$

$$c_i = 0, \ \ i = 1, \dots m, \quad c_{2m+1} = 1$$

$$x^T = [w_1, w_2, \dots, w_m, s]$$

In the dual case we choose the dual variable $Z \geq 0$ as

$$Z = \begin{bmatrix} z_1 \\ z_2 \end{bmatrix} \cdot [z_1^T \quad z_2^T] \tag{9}$$

where $z_1$ and $z_2$, are $m \times 1$ and $(m+1) \times 1$ column vectors, respectively. Obviously (9) choice of $Z$ implies that it is positive definite.

From the complementary slackness condition (4) we have

$$(sI - W_2)z_1 = 0 \tag{10-a}$$

$$(sI + W_1 - \boldsymbol{v}\boldsymbol{v}^T)z_2 = 0 \tag{10-b}$$

Multiplying both sides of equation (10-b) by $\boldsymbol{v}\boldsymbol{v}^T$ we have $s(\boldsymbol{v}\boldsymbol{v}^T z_2) = 0$ which implies that

$$\boldsymbol{v}^T z_2 = 0 \tag{11}$$

Using the constraints $Tr[F_i Z] = c_i$ we have

$$z_1^T z_1 + z_2^T z_2 = 1 \tag{12-a}$$

$$(\boldsymbol{\beta}_i^T z_1)^2 = (\boldsymbol{\alpha}_i^T z_2)^2 \quad \textbf{for} \quad i = 1, \dots, m \tag{12-b}$$

To have the strong duality we set $c^T x + Tr[F_0 Z] = 0$, hence we have

$$z_1^T z_1 - z_2^T z_2 = s$$

Considering the linear independence of $\boldsymbol{\alpha}_i$ and $\boldsymbol{\beta}_i$ for $i = 1, \dots, m$, we can expand $z_1$ and $z_2$ in terms of $\boldsymbol{\beta}_i$ and $\boldsymbol{\alpha}_i$as

$$z_1 = \sum_{i=1}^{m} a_i \boldsymbol{\beta}_i \tag{13-a}$$

$$z_2 = \sum_{i=1}^{m} a_i' \boldsymbol{\alpha}_i \tag{13-b}$$

with the coordinates $a_i, a_i^{'}$ for $i = 1, \dots, m$ to be determined.

Using (8) and the expansions (13), while considering (11), by comparing the coefficients of $\boldsymbol{\alpha}_i$ and $\boldsymbol{\beta}_i$ for $i = 1, \dots, m$ in the slackness conditions (10), we have

$$(-s+1)a_i = w_i \boldsymbol{\beta}_i^T z_1 \quad \textbf{for} \quad i = 1, \dots, m \tag{14-a}$$

$$(s+1)a_i' = w_i \boldsymbol{\alpha}_i^T z_2 \quad \textbf{for} \quad i = 1, \dots, m \tag{14-b}$$

Considering (12-b), we obtain

$$(-s+1)^2 a_i^2 = (s+1)^2 a_i'^2 \quad for \quad i = 1, \dots, m \tag{15}$$

or equivalently

$$\frac{a_i^2}{a_j^2} = \frac{a_i'^2}{a_j'^2} \tag{16}$$

for $\forall i,j = [1, m_1]$ and for $\boldsymbol{\beta}_i^T z_1$ and $\boldsymbol{\alpha}_i^T z_2$ $i = 1, \ldots m$, we have

$$\boldsymbol{\beta}_i^T z_1 = \sum_{j=1}^{m} a_j G_{i,j} \tag{17-a}$$

$$\boldsymbol{\alpha}_i^T z_2 = \sum_{j=1}^{m} a_j' G_{i,j}' \tag{17-b}$$

where $G$ and $G'$ are the gram matrices, defined as $G_{i,j} = \boldsymbol{\beta}_i^T \boldsymbol{\beta}_j$ and $G_{i,j}' = \boldsymbol{\alpha}_i^T \boldsymbol{\alpha}_j$, respectively, or equivalently

$$G = \begin{bmatrix} 1+nk & -1 & & \cdots & 0 \\ -1 & 2 & -1 & & \vdots \\ & -1 & \ddots & \ddots & 0 \\ \vdots & & \ddots & 2 & -1 \\ 0 & \cdots & 0 & -1 & 1+k \end{bmatrix}, \quad G' = \begin{bmatrix} 1 & -1 & & \cdots & 0 \\ -1 & 2 & -1 & & \vdots \\ & -1 & \ddots & \ddots & 0 \\ \vdots & & \ddots & 2 & -1 \\ 0 & \cdots & 0 & -1 & 1+k \end{bmatrix}$$

Substituting (17) in (14) we have

$$(-s + 1 - (1+nk)w_1)a_1 = -w_1 a_2 \tag{18-a}$$

$$(-s + 1 - 2w_i)a_i = -w_i(a_{i-1} + a_{i+1}) \ \ \textbf{for} \ \ i = 2, \ldots, m-1 \tag{18-b}$$

$$(-s + 1 - (1+k)w_m)a_m = -w_m a_{m-1} \tag{18-c}$$

and

$$(s + 1 - w_1)a_1' = -w_1 a_2' \tag{19-a}$$

$$(s + 1 - 2w_i)a_i' = -w_i(a_{i-1}' + a_{i+1}') \ \ \textbf{for} \ \ i = 2, \ldots, m \tag{19-b}$$

$$(s + 1 - (1+k)w_m)a_m' = -w_m a_{m-1}' \tag{19-c}$$

Now we can determine $s$ (*SLEM*), the optimal weights and the coordinates $a_i$ and $a_i'$, in an inductive manner as follows:

In the first stage, from comparing equations (18-c) and (19-c) and considering the relation (16), we can conclude that

$$(-s + 1 - (1+k)w_m)^2 = (s + 1 - (1+k)w_m)^2$$

which results in

$$w_m = 1/(1+k) \tag{20}$$

and $s = 0$, where the latter is not acceptable. Assuming $s = \cos(\theta)$ and substituting (20) in (18-c) and (19-c), we have

$$a_{m-1} = \frac{\sin(2\theta)}{\sin(\theta)} a_m,$$

$$a_{m-1}' = \frac{\sin(2(\pi - \theta))}{\sin(\pi - \theta)} a_m'.$$

Continuing the above procedure inductively, up to $i - 1$ stages, and assuming

$$a_{m-j} = \frac{\sin\big((j+1)\theta\big)}{\sin(\theta)} a_m, \qquad 0 \le \forall j \le i$$

and

$$a'_{m-j} = \frac{\sin\big((j+1)(\pi-\theta)\big)}{\sin(\pi-\theta)} a'_m \qquad 0 \le \forall j \le i$$

for the $i$-th stage, we get the following equations from comparison of equations (18-b) and (19-b),

$$\left((-s+1-2w_i)\frac{\sin\big((m-i+1)\theta\big)}{\sin(\theta)} + w_i\frac{\sin\big((m-i+2)\theta\big)}{\sin(\theta)}\right) a_m = -w_i a_{i+1} \tag{21-a}$$

$$\left((s+1-2w_i)\frac{\sin\big((m-i+1)(\pi-\theta)\big)}{\sin(\pi-\theta)} + w_i\frac{\sin\big((m-i+2)(\pi-\theta)\big)}{\sin(\pi-\theta)}\right) a'_m = -w_i a'_{i+1} \tag{21-b}$$

while considering relation (16) we can conclude that

$$\Big((-s+1-2w_i)\sin\big((m-i+1)\theta\big) + w_i\sin\big((m-i+2)\theta\big)\Big)^2$$

$$= \Big((s+1-2w_i)\sin\big((m-i+1)(\pi-\theta)\big) + w_i\sin\big((m-i+2)(\pi-\theta)\big)\Big)^2$$

which results in

$$w_i = \frac{1}{2} \tag{22}$$

Substituting $w_i = 1/2$ in (21), we have

$$a_{i+1} = \frac{\sin\big((m-i)\theta\big)}{\sin(\theta)} a_m \tag{23-a}$$

$$a'_{i+1} = \frac{\sin\big((m-i)(\pi-\theta)\big)}{\sin(\pi-\theta)} a'_m \tag{23-b}$$

the results in (22) and (23) are true for $i = 2, \dots, m-1$, and in the last stage, from equations (18-a) and (19-a) and using relation (16) and (23), we can conclude that

$$w_1 = \frac{2}{2+nk}$$

and $\theta$ has to satisfy the following relation

$$(nk-2)\big(\sin\big((m-1)\theta\big) + \sin\big((m+1)\theta\big)\big) + 2nk\sin(m\theta) = 0 \tag{24}$$

where *SLEM* equals $\cos(\theta)$ for the smallest root of (24) between 0 and $\pi$

Also one should notice that necessary and sufficient conditions for the convergence of weight matrix are satisfied, since all roots of $s$ which are the eigenvalues of $W$ are strictly less that one in magnitude, and one is a simple eigenvalue of $W$ associated with the eigenvector $\mathbf{1}$, where this happens due to positivity of optimal weights [6].

*B. Complete Cored Symmetric (CCS) Petal Network*

Here we consider a CCS Petal network of order $n, m, k$ with the undirected associated connectivity graph $\mathcal{G} = (\mathcal{V}, \mathcal{E})$ consisting of $|\mathcal{V}| = n(k(m-1)+2)$ nodes and $|\mathcal{E}| = \frac{n(n-1)}{2} + nk(m-1)$ edges, where the set of nodes is denoted by $\mathcal{V} = \{(i,j,q)\}$ where $i, j$ and $k$ vary from 1 to $m+1$, 1 to $n$ and 1 to $k$ respectively. (see Fig.3 for $n = 3, m = 4, k = 3$).

Automorphism of CCS Petal network is $S_k$ permutation of nodes of each leaf on $i$-th distance from central node for $i = 1, \dots, m$ and $S_n$ permutation of leaves hence according to subsection II-B it has $m+2$ class of edge orbits and it suffices to consider just $m+1$ weights $w_0, w_1, w_2, \dots, w_m$ (as labeled in Fig. 3. for $n = 3, m = 4, k = 3$), and consequently the weight matrix for the network can be defined as

$$W_{(i,j,q),(\mu,\rho,\eta)} = \begin{cases} w_0 \ \textbf{for} \ i = \mu = 1, \ j, \rho = 1, \dots, n, \ q = \eta = 1 \\ w_1 \ \textbf{for} \ i = 1, \mu = 2, \ j = \rho = 1, \dots, n, \ q = 1, \ \eta = 1, \dots, k \\ w_i \ \textbf{for} \ i = 2, \dots, m-1, \ \mu = i+1, \ j = \rho = 1, \dots, n, \ q = \eta = 1, \dots, k \\ w_m \ \textbf{for} \ i = m, \ \mu = m+1, \ j = \rho = 1, \dots, n, \ \eta = 1, \ q = 1, \dots, k \\ 1 - (n-1)w_0 - kw_1 \ \textbf{for} \ i = \mu = 1, \ j = \rho = 1, \dots, n, \ q = \eta = 1 \\ 1 - w_{i-1} - w_i \ \textbf{for} \ i = \mu = 2, \dots, m, \ j = \rho = 1, \dots, n, \ q = \eta = 1, \dots, k \\ 1 - kw_m \ \textbf{for} \ i = \mu = m+1 \ j = \rho = 1, \dots, n, \ q = 1, \dots, k, \ \eta = 1 \end{cases}$$

We associate with the node $(i,,j,q)$ the $|\mathcal{V}| \times 1$ column vector $e_{i,j,q} = e_i \otimes e_j \otimes e_q$ for $\{i,j,q\} = \{i = 1, \dots, m+1, \ j = 1, \dots, n, \ q = 1, \dots, k\}$ where $e_i, e_j$ and $e_q$ are $(m+1) \times 1$, $n \times 1$ and $k \times 1$ column vectors with one in the $i$-th, $j$-th and $q$-th position respectively and zero elsewhere. Introducing the new basis

$$\varphi_{i,\mu_1,\mu_2} = \frac{1}{\sqrt{nk}} \sum_{j=1}^{n} \omega_1^{(j-1)\mu_1} \sum_{q=1}^{k} \omega_2^{(q-1)\mu_2} e_{i,j,q} \quad \textbf{for} \ i = 1, \dots, m+1, \ \mu_1 = 0, \dots, n-1, \ \mu_2 = 0, \dots, k-1$$

with $\omega_1 = e^{j\frac{2\pi}{n}}$ and $\omega_2 = e^{j\frac{2\pi}{k}}$ and using Stratification method [7, 11, 12, 13@4th paper] the weight matrix $W$ for CCS Petal network in the new basis takes the block diagonal form with diagonal blocks $W_1, W_2$ and $W_3$ defined as:

$$W_1 = \begin{bmatrix} 1 - kw_1 & \sqrt{k}w_1 & 0 & & & \\ \sqrt{k}w_1 & 1 - w_1 - w_2 & w_2 & \ddots & & \\ 0 & w_2 & 1 - w_2 - w_3 & \ddots & 0 & \\ & \ddots & \ddots & \ddots & w_{m-1} & 0 \\ & & 0 & w_{m-1} & 1 - w_{m-1} - w_m & \sqrt{k}w_m \\ & & & 0 & \sqrt{k}w_m & 1 - kw_m \end{bmatrix} \qquad \text{(25-a)}$$

$$W_2 = \begin{bmatrix} 1 - kw_1 - nw_0 & \sqrt{k}w_1 & 0 & & & \\ \sqrt{k}w_1 & 1 - w_1 - w_2 & w_2 & \ddots & & \\ 0 & w_2 & 1 - w_2 - w_3 & \ddots & 0 & \\ & \ddots & \ddots & \ddots & w_{m-1} & 0 \\ & & 0 & w_{m-1} & 1 - w_{m-1} - w_m & \sqrt{k}w_m \\ & & & 0 & \sqrt{k}w_m & 1 - kw_m \end{bmatrix} \qquad \text{(25-b)}$$

$$W_3 = \begin{bmatrix} 1-w_1-w_2 & w_2 & 0 & & \\ w_2 & 1-w_2-w_3 & w_3 & \ddots & \\ 0 & w_3 & \ddots & \ddots & \\ & \ddots & \ddots & 1-w_{m-2}-w_{m-1} & w_{m-1} \\ & & & w_{m-1} & 1-w_{m-1}-w_m \end{bmatrix} \tag{25-c}$$

Considering the fact that $W_3$ is a submatrix of $W_1$ and $W_2$, respectively and using *Cauchy Interlacing Theorem*, we can state the following corollary for the eigenvalues of $W_1, W_2$ and $W_3$,

In the case of $n = 1$, after stratification the weight matrix $W_2$ does not exist and consequently Cauchy interlacing theorem will not be true thus the followings are true for $n \geq 2$.

*Corollary 2*,

For $W_1, W_2$ and $W_3$ given in (25), theorem 1 implies the following relations between the eigenvalues of $W_1, W_2$ and $W_3$

$$\lambda_{m+1}(W_1) \leq \lambda_m(W_1) \leq \lambda_{m-1}(W_3) \leq \cdots \leq \lambda_2(W_1) \leq \lambda_1(W_3) \leq \lambda_1(W_1) \tag{26-a}$$

$$\lambda_{m+1}(W_2) \leq \lambda_m(W_2) \leq \lambda_{m-1}(W_3) \leq \cdots \leq \lambda_2(W_2) \leq \lambda_1(W_3) \leq \lambda_1(W_2) \tag{26-b}$$

Considering the relation $W_1 = W_2 + nw_0e_1e_1^T$ between the matrices $W_1$ and $W_2$ and using the *Courant-Weyl inequalities* theorem,

*Theorem 2* (*The Courant-Weyl inequalities*) [16]:

Let $A$ and $B$ be Hermitean Matrices of order $n$, and let $1 \leq i,j \leq n$.

*(i)* If $i + j \leq n$ then $\lambda_{i+j-1}(A + B) \leq \lambda_i(A) + \lambda_j(B)$,

*(ii)* If $i + j - n \geq 1$ then $\lambda_i(A) + \lambda_j(B) \leq \lambda_{i+j-n}(A + B)$,

*(iii)* If $B$ is positive semidefinite, then $\lambda_i(A + B) \geq \lambda_i(A)$.

we state the following corollary for the eigenvalues of $W_1$ and $W_2$,

*Corollary 3*,

For $w_0 > 0$ and $W_1, W_2$ given in (25-a) and (25-b), respectively, theorem 2 implies the following interlacing relations between the eigenvalues of $W_1$ and $W_2$

$$\lambda_{|\mathcal{V}|}(W) = \lambda_{m+1}(W_2) \leq \lambda_{m+1}(W_1) \leq \lambda_m(W_2) \leq \cdots \leq \lambda_2(W_1) \leq \lambda_1(W_2) \leq \lambda_1(W_1) = 1$$

It is obvious from corollary 2 and 3 that $\lambda_2(W)$ and $\lambda_{|\mathcal{V}|}(W)$ are amongst the eigenvalues of $W_2$.

Based on corollary 2 and 3 and subsection II-A, one can express FDC problem for CCS Patel network in the form of semidefinite programming as:

$$\begin{aligned} \min \quad & s \\ s.t. \quad & -sI \leq W_2 \leq sI \end{aligned} \tag{27}$$

$W_2$ can be written as

$$W_2 = I_{m+1} - \sum_{i=0}^{m} w_i \boldsymbol{\alpha}_i \boldsymbol{\alpha}_i^T \tag{28}$$

where $\boldsymbol{\alpha}_i$ for $i = 0, \dots, m$ are $(m+1) \times 1$ column vectors, respectively defined as:

$$\boldsymbol{\alpha}_0(j) = \begin{cases} -\sqrt{n} & j = 1 \\ 0 & Otherwise \end{cases}$$

$$\boldsymbol{\alpha}_1(j) = \begin{cases} \sqrt{k} & j = 1 \\ -1 & j = 2 \\ 0 & Otherwise \end{cases}$$

$$\boldsymbol{\alpha}_i(j) = \begin{cases} 1 & j = i \\ -1 & j = i+1 \\ 0 & Otherwise \end{cases} \quad \textbf{for} \quad i = 2, \dots, m-1,$$

$$\boldsymbol{\alpha}_m(j) = \begin{cases} 1 & j = m \\ -\sqrt{k} & j = m+1 \\ 0 & Otherwise \end{cases}$$

In order to formulate problem (27) in the form of standard semidefinite programming described in section II-C, we define $F_i, c_i$ and $x$ as below:

$$F_0 = \begin{bmatrix} -I_{m+1} & 0 \\ 0 & I_{m+1} \end{bmatrix}$$

$$F_i = \begin{bmatrix} \boldsymbol{\alpha}_{i-1}\boldsymbol{\alpha}_{i-1}^T & 0 \\ 0 & -\boldsymbol{\alpha}_{i-1}\boldsymbol{\alpha}_{i-1}^T \end{bmatrix} \quad \textbf{for} \quad i = 1, \dots, m+1,$$

$$F_{m+2} = I_{2m+2}$$

$$c_i = 0, \ \ i = 1, \dots m+1, \quad c_{m+2} = 1$$

$$x^T = [w_0, w_1, w_2, \dots, w_m, s]$$

In the dual case we choose the dual variable $Z \geq 0$ as

$$Z = \begin{bmatrix} z_1 \\ z_2 \end{bmatrix} \cdot \begin{bmatrix} z_1^T & z_2^T \end{bmatrix} \tag{29}$$

where $z_1$ and $z_2$, are $(m+1) \times 1$ and $(m+1) \times 1$ column vectors, respectively. Obviously (29) choice of $Z$ implies that it is positive definite.

From the complementary slackness condition (4) we have

$$(sI - W_2)z_1 = 0 \tag{30-a}$$

$$(sI + W_2)z_2 = 0 \tag{30-b}$$

Using the constraints $Tr[F_i Z] = c_i$ we have

$$z_1^T z_1 + z_2^T z_2 = 1 \tag{31-a}$$

$$(\boldsymbol{\alpha}_i^T z_1)^2 = (\boldsymbol{\alpha}_i^T z_2)^2 \quad \textbf{for} \quad i = 0, \dots, m \tag{31-b}$$

To have the strong duality we set $c^T x + Tr[F_0 Z] = 0$, hence we have

$$z_1^T z_1 - z_2^T z_2 = s$$

Considering the linear independence of $\boldsymbol{\alpha}_i$ for $i = 0, \dots, m$, we can expand $z_1$ and $z_2$ in terms of $\boldsymbol{\alpha}_i$as

$$z_1 = \sum_{i=0}^{m} a_i \boldsymbol{\alpha}_i \tag{32-a}$$

$$z_2 = \sum_{i=0}^{m} a_i' \boldsymbol{\alpha}_i \tag{32-b}$$

with the coordinates $a_i, a_i'$ for $i = 0, \dots, m$ to be determined.

Using (28) and the expansions (32), while considering (11), by comparing the coefficients of $\boldsymbol{\alpha}_i$ $i = 0, \dots, m$ in the slackness conditions (30), we have

$$(-s+1)a_i = w_i \boldsymbol{\alpha}_i^T z_1 \quad \textbf{for} \quad i = 0, \dots, m \tag{33-a}$$

$$(s+1)a_i' = w_i \boldsymbol{\alpha}_i^T z_2 \quad \textbf{for} \quad i = 0, \dots, m \tag{33-b}$$

Considering (31-b), we obtain

$$(-s+1)^2 a_i^2 = (s+1)^2 a_i'^2 \quad \textbf{for} \quad i = 0, \dots, m \tag{34}$$

or equivalently

$$\frac{a_i^2}{a_j^2} = \frac{a_i'^2}{a_j'^2} \tag{35}$$

for $\forall i, j = [0, m]$ and for $\boldsymbol{\alpha}_i^T z_1$ and $\boldsymbol{\alpha}_i^T z_2$ $i = 0, \dots m$, we have

$$\boldsymbol{\alpha}_i^T z_1 = \sum_{j=0}^{m} a_j G_{i,j} \tag{36-a}$$

$$\boldsymbol{\alpha}_i^T z_2 = \sum_{j=0}^{m} a_j' G_{i,j} \tag{36-b}$$

where $G$ is the gram matrices, defined as $G_{i,j} = \boldsymbol{\alpha}_i^T \boldsymbol{\alpha}_j$, or equivalently

$$G = \begin{bmatrix} n & -\sqrt{kn} & & \cdots & & 0 \\ -\sqrt{kn} & k+1 & -1 & & & \vdots \\ & -1 & 2 & \ddots & & \\ \vdots & & \ddots & \ddots & -1 & 0 \\ & & & -1 & 2 & -1 \\ 0 & \cdots & & 0 & -1 & 1+k \end{bmatrix},$$

Substituting (36) in (33) we have

$$(-s+1-nw_0)\hat{a}_0 = -knw_0 a_1 \tag{37-a}$$

$$(-s+1-(k+1)w_1)a_1 = -w_1(\hat{a}_0 + a_2) \tag{37-b}$$

$$(-s+1-2w_i)a_i = -w_i(a_{i-1} + a_{i+1}) \quad \textbf{for} \quad i = 2,\dots,m-1 \tag{37-c}$$

$$(-s+1-(1+k)w_m)a_m = -w_m a_{m-1} \tag{37-d}$$

and

$$(s+1-nw_0)\hat{a}'_0 = -knw_0 a'_1 \tag{38-a}$$

$$(s+1-(k+1)w_1)a'_1 = -w_1(\hat{a}'_0 + a'_2) \tag{38-b}$$

$$(s+1-2w_i)a'_i = -w_i(a'_{i-1} + a'_{i+1}) \quad \textbf{for} \quad i = 2,\dots,m \tag{38-c}$$

$$(s+1-(1+k)w_m)a'_m = -w_m a'_{m-1} \tag{38-d}$$

where $\hat{a}_0 = \sqrt{kn}a_0$ and $\hat{a}'_0 = \sqrt{kn}a'_0$. Now we can determine $s$ (*SLEM*), the optimal weights and the coordinates $a_i$ and $a'_i$, in an inductive manner as follows:

In the first stage, from comparing equations (37-a) and (38-a) and considering the relation (35), we can conclude that

$$(-s+1-nw_0)^2 = (s+1-nw_0)^2$$

which results in $w_0 = 1/n$ and $s = 0$, where the latter is not acceptable. Substituting $w_0 = 1/n$ in (37-a) and (38-a), we have

$$a_1 = \frac{s}{k}\hat{a}_0,$$

$$a'_1 = -\frac{s}{k}\hat{a}'_0$$

In the second stage, from comparing equations (37-b) and (38-b) and considering the relation (35), we can conclude that

$$\left((-s+1-(k+1)w_1)\frac{s}{k} + w_1\right)^2 = \left(-(s+1-(k+1)w_1)\frac{s}{k} + w_1\right)^2$$

which results in $w_1 = 1/(k+1)$, where the latter is not acceptable. Substituting $w_1 = 1/(k+1)$ in (37-b) and (38-b), we have

$$a_2 = \left(\frac{k+1}{k}s^2 - 1\right)\hat{a}_0$$

$$a'_2 = \left(\frac{k+1}{k}s^2 - 1\right)\hat{a}'_0$$

Continuing the above procedure inductively, up to $i-1$ stages, and assuming

$$a_j = f_j(s)\hat{a}_0, \qquad 1 \le \forall j \le i$$

and

$$a'_j = f_j(-s)\hat{a}'_0 \qquad 1 \le \forall j \le i$$

where $f_i(s)$ is an even (odd) polynomial in terms of $s$ for even (odd) values of $i$, for the $k$-th stage, by comparing equations (37-c) and (38-c), we get the following equations

$$\big((-s+1-2w_i)f_i(s)+w_i f_{i-1}(s)\big)\hat{a}_0=-w_i a_{i+1} \tag{39-a}$$

$$\big((s+1-2w_i)f_i(-s)+w_i f_{i-1}(-s)\big)\hat{a}'_0=-w_i a'_{i+1} \tag{39-b}$$

and considering relation (35) we can conclude that

$$\big((-s+1-2w_i)f_i(s)+w_i f_{i-1}(s)\big)^2=\big((-s+1-2w_i)f_i(s)+w_i f_{i-1}(s)\big)^2$$

which results in

$$w_i=\frac{1}{2} \tag{40}$$

Substituting $w_i=1/2$ in (39) we have

$$\big(2sf_i(s)-f_{i-1}(s)\big)\hat{a}_0=a_{i+1} \tag{41-a}$$

$$\big(-2sf_i(-s)-f_{i-1}(-s)\big)\hat{a}'_0=a'_{i+1} \tag{41-b}$$

where (40), and (41) hold true for $i=1,\dots,m-1$, and in the $m$-th stage, from equations (37-d) and (38-d) and using relation (35), we can conclude that

$$w_m=1/(k+1) \tag{42}$$

$$a_m=\big(2sf_{m-1}(s)-f_{m-2}(s)\big)\hat{a}_0 \tag{43-a}$$

$$a'_m=\big(-2sf_{m-1}(-s)-f_{m-2}(-s)\big)\hat{a}'_0 \tag{43-b}$$

and $s$ has to satisfy following equation

$$(2(k+1)s^2-1)f_{m-1}(s)=(k+1)sf_{m-2}(s) \tag{44}$$

The polynomials $f_i(s)$ can be obtained inductively as follows:

$$f_1(s)=\frac{s}{k}$$

$$f_2(s)=\frac{(k+1)}{k}s^2-1$$

$$f_i(s)=2sf_{i-1}(s)-f_{i-2}(s)\ \ \textbf{for}\ \ i=3,\dots,m-1$$

*C. Asymmetric $G_{m_1,m_2}$ Leaf*

Here we consider a symmetric Petal network with $\mathrm{G}_{m,m'}$ leaves where every node at $i$-th distance from the central node have $k_i, i=1,\dots,m+m'$ children, except the leaf nodes. The network's undirected associated connectivity graph $\mathcal{G}=(\mathcal{V},\mathcal{E})$ consists of $|\mathcal{V}|=1+n\big(\sum_{i=1}^{m+m'}\prod_{j=1}^{i}k_j+1\big)$ nodes and $|\mathcal{E}|=n\big(\sum_{i=1}^{m+m'}\prod_{j=1}^{i}k_j+1\big)$ edges, where the set of nodes on $i$-th distance of central node is denoted by $\mathcal{V} = \{(j, q_1, q_2, \dots, q_i)\}$ where $i, j$ and $q_i$ vary from 1 to $m + m'$, 1 to $n$ and 1 to $k_i$ respectively. (See Fig.5 for $m = 3, m' = 2, k_1 = k_2 = k_3 = k_1' = 2, k_2' = 4$) and we indicate the central node by (1).

Automorphism of symmetric Petal network with $G_{m,m'}$ leaves is $S_{k_i}$ permutation of nodes of each leaf on $i$-th distance from central node for $i = 1, \dots, m + m'$ and $S_n$ permutation of leaves hence according to subsection II-B it has $m + m' + 1$ class of edge orbits and it suffices to consider just $m + m'$ weights $w_1, w_2, \dots, w_{m+m'}$ (as labeled in Fig. 5. for $m = 3, m' = 2$,), and consequently the weight matrix for the network can be defined as

$$W_{i,j} = \begin{cases} w_\mu \ \ \textbf{for} \ \ i = (r, q_1, \dots, q_{\mu-1}), \ j = (r, q_1, \dots, q_{\mu-1}, q_\mu), \ r = 1, \dots, n, \ q_L = 1, \dots, k_L, \ L = 1, \dots, \mu, \ \mu = 1, \dots, m + m' \\ 1 - nk_1 w_1 \ \ \textbf{for} \ \ i = j = (1), \\ 1 - w_{m+m'} \ \ \textbf{for} \ \ i = j = (r, q_1, \dots, q_{m+m'}), \ q_L = 1, \dots, k_L, \ L = 1, \dots, m + m', \\ 1 - k_\mu w_{\mu+1} - w_\mu \ \ \textbf{for} \ \ i = j = (1, q_1, \dots, q_\mu), \ q_L = 1, \dots, k_L, \ L = 1, \dots, \mu, \ \mu = 1, \dots, m + m' - 1, \end{cases}$$

We associate with the node $(r, q_1, \dots, q_\mu)$ the $|\mathcal{V}| \times 1$ column vector $e_{r,q_1,\dots,q_{\mu-1}} = e_r \otimes e_{q_1} \otimes \dots \otimes e_{q_\mu}$ for $\{r, q_1, \dots, q_\mu\} = \{\mu = 1, \dots, m + m', \ r = 1, \dots, n, \ q_L = 1, \dots, k_L, \ L = 1, \dots, \mu\}$ where $e_r, e_{q_1}, \dots, e_{q_{\mu-1}}$ and $e_{q_\mu}$ are $n \times 1, q_1 \times 1, \dots, q_{\mu-1} \times 1$ and $q_\mu \times 1$ column vectors with one in the $r$-th, $q_1$-th, $\dots, q_{\mu-1}$-th and $q_\mu$-th position respectively and zero elsewhere. Introducing the new basis $\varphi_1 = (1)$ and

$$\varphi_\mu = \frac{1}{\sqrt{n \prod_{i=1}^{\mu} k_i}} \sum_{j=1}^{n} \omega_0^{(r-1)\rho_0} \sum_{q_1=1}^{k_1} \omega_1^{(q_1-1)\rho_1} \dots \sum_{q_\mu=1}^{k_\mu} \omega_\mu^{(q_\mu-1)\rho_\mu} e_{(r,q_1,\dots,q_\mu)} \ \ \textbf{for} \ \ \mu = 1, \dots, m + m', \ \rho_0 = 0, \dots, n - 1, \ \rho_L = 0, \dots, k_L - 1, \ L = 1, \dots, \mu$$

with $\omega_0 = e^{j\frac{2\pi}{n}}$ and $\omega_L = e^{j\frac{2\pi}{k_L}}$ for $L = 1, \dots, \mu$ and using Stratification method [7, 11, 12, 13@4[th] paper] the weight matrix $W$ for CCS Petal network in the new basis takes the block diagonal form with diagonal blocks $W_1, W_2, \dots, W_{\left\lfloor\frac{m+m'+1}{2}\right\rfloor+2}$ defined as:

$$W_1 = \begin{bmatrix} 1 - nk_1 w_1 & \sqrt{nk_1} w_1 & 0 & & & & & & \\ \sqrt{nk_1} w_1 & 1 - w_1 - k_2 w_2 & \sqrt{k_2} w_2 & \ddots & & & & & \\ 0 & \sqrt{k_2} w_2 & 1 - w_2 - k_3 w_3 & \ddots & 0 & & & & \\ & \ddots & \ddots & \ddots & \sqrt{k_m} w_m & 0 & & & \\ & & 0 & \sqrt{k_m} w_m & 1 - w_m - w_{m+1} & \sqrt{k_{m+1}} w_{m+1} & & \ddots & \\ & & & 0 & \sqrt{k_{m+1}} w_{m+1} & \ddots & & \ddots & 0 \\ & & & & \ddots & \ddots & & 1 - k_{m+m'-1} w_{m+m'-1} - w_{m+m'} & \sqrt{k_{m+m'}} w_{m+m'} \\ & & & & & 0 & & \sqrt{k_{m+m'}} w_{m+m'} & 1 - k_{m+m'} w_{m+m'} \end{bmatrix} \quad \text{(45-a)}$$

$$W_2 = \begin{bmatrix} 1 - w_1 - k_2 w_2 & \sqrt{k_2} w_2 & \ddots & & & & & \\ \sqrt{k_2} w_2 & 1 - w_2 - k_3 w_3 & \ddots & 0 & & & & \\ \ddots & \ddots & \ddots & \sqrt{k_m} w_m & 0 & & & \\ & 0 & \sqrt{k_m} w_m & 1 - w_m - w_{m+1} & \sqrt{k_{m+1}} w_{m+1} & & \ddots & \\ & & 0 & \sqrt{k_{m+1}} w_{m+1} & \ddots & & \ddots & 0 \\ & & & \ddots & \ddots & & 1 - k_{m+m'-1} w_{m+m'-1} - w_{m+m'} & \sqrt{k_{m+m'}} w_{m+m'} \\ & & & & 0 & & \sqrt{k_{m+m'}} w_{m+m'} & 1 - k_{m+m'} w_{m+m'} \end{bmatrix} \quad \text{(45-b)}$$

$$W_3 = \begin{bmatrix} 1-w_1-k_2w_2 & \sqrt{k_2}w_2 & \ddots & & & & \\ \sqrt{k_2}w_2 & 1-w_2-k_3w_3 & \ddots & 0 & & & \\ \ddots & \ddots & \ddots & \sqrt{k_m}w_m & 0 & & \\ & 0 & \sqrt{k_m}w_m & 1-w_m-w_{m+1} & \sqrt{k_{m+1}}w_{m+1} & & \ddots \\ & & 0 & \sqrt{k_{m+1}}w_{m+1} & \ddots & & \ddots \\ & & & \ddots & \ddots & 1-k_{m+m'-1}w_{m+m'-1}-w_{m+m'} & \end{bmatrix} \quad \text{(45-c)}$$

and $W_i$ for $i = 4, \ldots, \left\lfloor\frac{m+m'+1}{2}\right\rfloor + 2$ obtains from deleting first and last row and column of $W_{i-1}$.

Considering the fact that $W_2$ is a submatrix of $W_1$ and using *Cauchy Interlacing Theorem*, we can state the following corollary for the eigenvalues of $W_1$ and $W_2$,

In the case of $n = 1$, after stratification the weight matrix $W_2$ does not exist and consequently Cauchy interlacing theorem will not be true thus the followings are true for $n \geq 2$.

*Corollary 4*,

For $W_1$ and $W_2$ given in (45), theorem 1 implies the following relations between the eigenvalues of $W_1$ and $W_2$

$$\lambda_{|\mathcal{V}|}(W) = \lambda_{m+m'+1}(W_1) \leq \lambda_{m+m'}(W_2) \leq \lambda_{m+m'}(W_1) \leq \cdots \leq \lambda_2(W_1) \leq \lambda_1(W_2) \leq \lambda_1(W_1) = 1$$

It is obvious from corollary 4 that $\lambda_2(W)$ and $\lambda_{|\mathcal{V}|}(W)$ are amongst the eigenvalues of $W_2$ and $W_1$, respectively.

Based on corollary 4 and subsection II-A, one can express FDC problem for the symmetric Patel network with $\mathrm{G}_{m,m'}$ leaves in the form of semidefinite programming as:

$$\begin{aligned} \min \quad & s \\ s.t. \quad & W_2 \leq sI \\ & -sI \leq W_1 - \boldsymbol{v}\boldsymbol{v}^T \end{aligned} \quad \text{(46)}$$

where $\boldsymbol{v}$ is a $(m+m'+1) \times 1$ column vector defined as:

$$\boldsymbol{v}(i) = \frac{1}{|\boldsymbol{v}|} \times \begin{cases} \sqrt{\prod_{j=m+1}^{m+m'} k_j} & \textbf{for} \quad i = 1 \\ \sqrt{\prod_{j=m+1}^{m+m'} k_j} \times \sqrt{\prod_{j=1}^{i-1} k_j} & \textbf{for} \quad i = 2, \ldots, m+1 \\ \sqrt{\prod_{j=i}^{m+m'} k_j} \times \sqrt{\prod_{j=1}^{m} k_j} & \textbf{for} \quad i = m+2, \ldots, m+m' \\ \sqrt{\prod_{j=1}^{m} k_j} & \textbf{for} \quad i = m+m'+1 \end{cases}$$

which is eigenvector of $W_1$ corresponding to the eigenvalue one. The matrices $W_1$ and $W_2$ can be written as

$$W_1 = I_{m+m'+1} - \sum_{i=1}^{m+m'} w_i \boldsymbol{\alpha}_i \boldsymbol{\alpha}_i^T \quad \text{(47-a)}$$

$$W_2 = I_{m+m'+1} - \sum_{i=1}^{m+m'} w_i \boldsymbol{\beta}_i \boldsymbol{\beta}_i^T \tag{47-b}$$

where $\boldsymbol{\alpha}_i$and $\boldsymbol{\beta}_i$ for $i = 1, \dots, m + m'$ are $(m + m') \times 1$ column vectors, respectively defined as:

$$\boldsymbol{\alpha}_1(j) = \begin{cases} \sqrt{k_1 n} & j = 1 \\ -1 & j = 2 \\ 0 & Otherwise \end{cases}$$

$$\boldsymbol{\alpha}_i(j) = \begin{cases} \sqrt{k_i} & j = i \\ -1 & j = i + 1 \\ 0 & Otherwise \end{cases} \quad \textbf{for} \quad i = 2, \dots, m$$

$$\boldsymbol{\alpha}_i(j) = \begin{cases} 1 & j = i \\ -\sqrt{k_i} & j = i + 1 \\ 0 & Otherwise \end{cases} \quad \textbf{for} \quad i = m + 1, \dots, m + m',$$

$$\boldsymbol{\beta}_1(j) = \begin{cases} -1 & j = 1 \\ 0 & Otherwise \end{cases}$$

$$\boldsymbol{\beta}_i(j) = \begin{cases} \sqrt{k_i} & j = i - 1 \\ -1 & j = i \\ 0 & Otherwise \end{cases} \quad \textbf{for} \quad i = 2, \dots, m$$

$$\boldsymbol{\beta}_i(j) = \begin{cases} 1 & j = i - 1 \\ -\sqrt{k_i} & j = i \\ 0 & Otherwise \end{cases} \quad \textbf{for} \quad i = m + 1, \dots, m + m',$$

In order to formulate problem (46) in the form of standard semidefinite programming described in section II-C, we define $F_i, c_i$ and $x$ as below:

$$F_0 = \begin{bmatrix} -I_{m+m'} & 0 \\ 0 & I_{m+m'+1} - \boldsymbol{v}\boldsymbol{v}^T \end{bmatrix}$$

$$F_i = \begin{bmatrix} \boldsymbol{\beta}_i \boldsymbol{\beta}_i^T & 0 \\ 0 & -\boldsymbol{\alpha}_i \boldsymbol{\alpha}_i^T \end{bmatrix} \quad \textbf{for} \quad i = 1, \dots, m + m',$$

$$F_{m+m'+1} = I_{2m+2m'+1}$$

$$c_i = 0, \;\; i = 1, \dots, m + m', \quad c_{m+m'+1} = 1$$

$$x^T = [w_1, w_2, \dots, w_{m+m'}, s]$$

In the dual case we choose the dual variable $Z \geq 0$ as

$$Z = \begin{bmatrix} z_1 \\ z_2 \end{bmatrix} \cdot \begin{bmatrix} z_1^T & z_2^T \end{bmatrix} \tag{48}$$

where $z_1$ and $z_2$, are $(m + m') \times 1$ and $(m + m' + 1) \times 1$ column vectors, respectively. Obviously (48) choice of $Z$ implies that it is positive definite.

From the complementary slackness condition (4) we have

$$(sI - W_2)z_1 = 0 \tag{49-a}$$

$$(sI + W_1 - \boldsymbol{v}\boldsymbol{v}^T)z_2 = 0 \tag{49-b}$$

Multiplying both sides of equation (49-b) by $\boldsymbol{v}\boldsymbol{v}^T$ we have $s(\boldsymbol{v}\boldsymbol{v}^T z_2) = 0$ which implies that

$$\boldsymbol{v}^T z_2 = 0 \tag{50}$$

Using the constraints $Tr[F_i Z] = c_i$ we have

$$z_1^T z_1 + z_2^T z_2 = 1 \tag{51-a}$$

$$(\boldsymbol{\beta}_i^T z_1)^2 = (\boldsymbol{\alpha}_i^T z_2)^2 \quad \textbf{for} \quad i = 1, \dots, m + m' \tag{51-b}$$

To have the strong duality we set $c^T x + Tr[F_0 Z] = 0$, hence we have

$$z_1^T z_1 - z_2^T z_2 = s$$

Considering the linear independence of $\boldsymbol{\alpha}_i$ and $\boldsymbol{\beta}_i$ for $i = 1, \dots, m + m'$, we can expand $z_1$ and $z_2$ in terms of $\boldsymbol{\alpha}_i$ and $\boldsymbol{\beta}_i$ as

$$z_1 = \sum_{i=1}^{m+m'} a_i \boldsymbol{\beta}_i \tag{52-a}$$

$$z_2 = \sum_{i=1}^{m+m'} a'_i \boldsymbol{\alpha}_i \tag{52-b}$$

with the coordinates $a_i, a'_i$ for $i = 1, \dots, m + m'$ to be determined.

Using (47) and the expansions (52), while considering (50), by comparing the coefficients of $\boldsymbol{\alpha}_i$ and $\boldsymbol{\beta}_i$ $i = 1, \dots, m + m'$ in the slackness conditions (49), we have

$$(-s + 1)a_i = w_i \boldsymbol{\beta}_i^T z_1 \quad \textbf{for} \quad i = 1, \dots, m + m' \tag{53-a}$$

$$(s + 1)a'_i = w_i \boldsymbol{\alpha}_i^T z_2 \quad \textbf{for} \quad i = 1, \dots, m + m' \tag{53-b}$$

Considering (51-b), we obtain

$$(-s + 1)^2 a_i^2 = (s + 1)^2 a_i'^2 \quad \textbf{for} \quad i = 1, \dots, m + m'$$

or equivalently

$$\frac{a_i^2}{a_j^2} = \frac{a_i'^2}{a_j'^2} \tag{54}$$

for $\forall i, j = [1, m + m']$ and for $\boldsymbol{\beta}_i^T z_1$ and $\boldsymbol{\alpha}_i^T z_2$ $i = 1, \dots, m + m'$, we have

$$\boldsymbol{\beta}_i^T z_1 = \sum_{j=1}^{m+m'} a_j G'_{i,j} \tag{55-a}$$

$$\boldsymbol{\alpha}_i^T z_2 = \sum_{j=0}^{m} a_j' G_{i,j} \tag{55-b}$$

where $G$ and $G'$ are the gram matrices, defined as $G_{i,j} = \boldsymbol{\alpha}_i^T \boldsymbol{\alpha}_j$ and $G'_{i,j} = \boldsymbol{\beta}_i^T \boldsymbol{\beta}_j$, or equivalently

$$G' = \begin{bmatrix}
1 & -\sqrt{k_2} & 0 & \cdots & & & & & & 0 \\
-\sqrt{k_2} & k_2+1 & -\sqrt{k_3} & 0 & & & & & & \vdots \\
0 & -\sqrt{k_3} & k_3+1 & \ddots & \ddots & & & & & \\
\vdots & 0 & \ddots & \ddots & -\sqrt{k_m} & 0 & & & & \\
 & & \ddots & -\sqrt{k_m} & k_m+1 & -1 & 0 & & & \\
 & & & 0 & -1 & 1+k_{m+1} & \sqrt{k_{m+1}} & \ddots & & \\
 & & & & 0 & \sqrt{k_{m+1}} & \ddots & \ddots & 0 & \\
 & & & & & \ddots & \ddots & 1+k_{m+m'-2} & -\sqrt{k_{m+m'-2}} & 0 \\
 & & & & & & 0 & -\sqrt{k_{m+m'-2}} & 1+k_{m+m'-1} & -\sqrt{k_{m+m'-1}} \\
0 & \cdots & & & & & & 0 & -\sqrt{k_{m+m'-1}} & 1+k_{m+m'}
\end{bmatrix},$$

$$G = \begin{bmatrix}
1+k_1 n & -\sqrt{k_2} & 0 & \cdots & & & & & & 0 \\
-\sqrt{k_2} & k_2+1 & -\sqrt{k_3} & 0 & & & & & & \vdots \\
0 & -\sqrt{k_3} & k_3+1 & \ddots & \ddots & & & & & \\
\vdots & 0 & \ddots & \ddots & -\sqrt{k_m} & 0 & & & & \\
 & & \ddots & -\sqrt{k_m} & k_m+1 & -1 & 0 & & & \\
 & & & 0 & -1 & 1+k_{m+1} & \sqrt{k_{m+1}} & \ddots & & \\
 & & & & 0 & \sqrt{k_{m+1}} & \ddots & \ddots & 0 & \\
 & & & & & \ddots & \ddots & 1+k_{m+m'-2} & -\sqrt{k_{m+m'-2}} & 0 \\
 & & & & & & 0 & -\sqrt{k_{m+m'-2}} & 1+k_{m+m'-1} & -\sqrt{k_{m+m'-1}} \\
0 & \cdots & & & & & & 0 & -\sqrt{k_{m+m'-1}} & 1+k_{m+m'}
\end{bmatrix},$$

Substituting (55) in (53) we have

$$(-s+1-w_1)a_1 = -w_1\sqrt{k_2}a_2 \tag{56-a}$$

$$\big(-s+1-w_i(1+k_i)\big)a_i = -w_i\big(\sqrt{k_i}a_{i-1} + \sqrt{k_{i+1}}a_{i+1}\big) \quad \textbf{for} \quad i = 2, \dots, m-1 \tag{56-b}$$

$$\big(-s+1-w_m(1+k_m)\big)a_m = -w_m\big(\sqrt{k_m}a_{m-1} + a_{m+1}\big) \tag{56-c}$$

$$\big(-s+1-w_{m+1}(1+k_{m+1})\big)a_{m+1} = -w_{m+1}\big(a_m + \sqrt{k_{m+1}}a_{m+1}\big) \tag{56-d}$$

$$\big(-s+1-w_i(1+k_i)\big)a_i = -w_i\big(\sqrt{k_{i-1}}a_{i-1} + \sqrt{k_i}a_{i+1}\big) \quad \textbf{for} \quad i = m+2, \dots, m+m'-1 \tag{56-e}$$

$$\big(-s+1-w_{m+m'}(1+k_{m+m'})\big)a_{m+m'} = -w_{m+m'}\sqrt{k_{m+m'-1}}a_{m+m'-1} \tag{56-f}$$

and

$$(s+1-(1+k_1 n)w_1)a'_1 = -w_1\sqrt{k_2}a'_2 \quad (57\text{-a})$$

$$\big(s+1-w_i(1+k_i)\big)a'_i = -w_i\big(\sqrt{k_i}a'_{i-1} + \sqrt{k_{i+1}}a'_{i+1}\big) \ \ \textbf{for} \ \ i = 2, \dots, m-1 \quad (57\text{-b})$$

$$\big(s+1-w_m(1+k_m)\big)a'_m = -w_m\big(\sqrt{k_m}a'_{m-1} + a'_{m+1}\big) \quad (57\text{-c})$$

$$\big(s+1-w_{m+1}(1+k_{m+1})\big)a'_{m+1} = -w_{m+1}\big(a'_m + \sqrt{k_{m+1}}a'_{m+1}\big) \quad (57\text{-d})$$

$$\big(-s+1-w_i(1+k_i)\big)a'_i = -w_i\big(\sqrt{k_{i-1}}a'_{i-1} + \sqrt{k_i}a'_{i+1}\big) \ \ \textbf{for} \ \ i = m+2, \dots, m+m'-1 \quad (57\text{-e})$$

$$\big(-s+1-w_{m+m'}(1+k_{m+m'})\big)a'_{m+m'} = -w_{m+m'}\sqrt{k_{m+m'-1}}a'_{m+m'-1} \quad (57\text{-f})$$

Continuing the same inductive procedure as in two previous subsections, the optimal weights are obtained as follows:

$$w_1 = \frac{2}{2+nk_1}$$

$$w_i = \frac{1}{1+k_i} \quad for \quad i = 2, \dots, m+m'$$

In the case of asymmetric $G_{m,m'}$ leaves with complete cored configuration, by employing the same method used for asymmetric $G_{m,m'}$ leaves with central node configuration, we can obtain the optimal weights as given in section III-D.

Symmetric $G_{m,k}$ leaves are asymmetric $G_{m,m'}$ leaves with $k_1 = k_2 = \cdots = k_{m+m'} = k$ where all obtained results for asymmetric $G_{m,m'}$ leaves holds true for symmetric $G_{m,k}$ leaves by replacing $k_1, k_2, \dots, k_{m+m'}$ by $k$.

## V. Conclusion

Fastest Distributed Consensus averaging Algorithm in sensor networks has received renewed interest recently, but Most of the methods proposed so far usually avoid the direct computation of optimal weights and deal with the FDC problem by numerical convex optimization methods.

Here in this work, we have solved FDC problem over symmetric petal network, with central node and complete cored configurations consisting of different kinds of leaves. It has been shown that the *SLEM* of networks with complete cored configuration do not change with number of leaves due to highway property of the central complete graph (in contrast with the bottleneck property of central node in symmetric Petal configuration). Also the obtained optimal weights do not depend on length of path graphs or height of tree graphs within the leaves.

Providing analytical solution of fastest distributed averaging problem over networks with more general topologies is the object of our future investigations.